\documentclass[aps,pr,twocolumn,superscriptaddress,amsmath,amssymb,showpacs,eqsecnum]{revtex4-1}
\usepackage{epsfig}
\usepackage{amsfonts}
\usepackage{braket}
\usepackage{epstopdf}
\usepackage{hyperref}
\usepackage{float}	
\usepackage{bibentry}
\usepackage[caption=false]{subfig}
\usepackage{mathrsfs}
\usepackage{graphicx}
\usepackage{dcolumn}
\usepackage{bm}
\usepackage{color}

\begin{document}
\title{Robust zero-energy bound states around a pair-density-wave vortex core in locally noncentrosymmetric superconductors}

\author{Yoichi Higashi}
\altaffiliation{Present Address: Department of Physics, Sungkyunkwan University, Suwon 16419, Korea}
\email{y.higashi@skku.edu}
\affiliation{Department of Mathematical Sciences, Osaka Prefecture University, 1-1 Gakuen-cho, Naka-ku, Sakai 599-8531, Japan}

\author{Yuki Nagai}
\affiliation{CCSE, Japan  Atomic Energy Agency, 178-4-4, Wakashiba, Kashiwa, Chiba 277-0871, Japan}

\author{Tomohiro Yoshida}
\altaffiliation{Present Address: Department of Physics, Gakushuin University, Tokyo 171-8588, Japan}
\affiliation{Graduate School of Science and Technology, Niigata University, Niigata 950-2181, Japan}

\author{Yusuke Masaki}
\affiliation{Department of Physics, The University of Tokyo, Tokyo 113-0033, Japan}

\author{Youichi Yanase}
\altaffiliation{Present Address: Department of Physics, Graduate School of Science, Kyoto University, Kyoto 606-8502, Japan}
\affiliation{Department of Physics, Niigata University, Niigata 950-2181, Japan}

\date{\today}
\begin{abstract}
We numerically investigate the electronic structures around a vortex core in a bilayer superconducting system, with
$s$-wave pairing, Rashba spin-orbit coupling and Zeeman magnetic field, with use of the quasiclassical Green's function method.
The Bardeen-Cooper-Schrieffer (BCS) phase and the so-called pair-density wave (PDW) phase appear
in the temperature-magnetic-field phase diagram 
in a bulk uniform system [Phys. Rev. B {\bf86}, 134514 (2012)].
In the low magnetic field perpendicular to the layers,
the zero-energy vortex bound states in the BCS phase are split by the Zeeman magnetic field.
On the other hand,
the PDW state appears in the high magnetic field,
and sign of the order parameter is opposite between the layers.
We find that the vortex core suddenly shrinks and the zero-energy bound states appear
by increasing the magnetic field through the BCS-PDW transition.
We discuss the origin of the change in vortex core structure between the BCS and PDW states
by clarifying the relation between the vortex bound states and
the bulk energy spectra.
In the high magnetic field region,
the PDW state and vortex bound states are protected by the spin-orbit coupling.
These characteristic behaviors in the PDW state can be observed by scanning tunneling microscopy/spectroscopy.
\end{abstract}

\maketitle
\section{Introduction}
Exploring unconventional or exotic superconducting phase has been attached great interest
and been one of the streams of the superconductivity research
\cite{RevModPhys.63.239,doi:10.1143/JPSJ.81.011009,sigrist-bauer2012},
since
the Cooper pair wave functions have internal degrees of freedom
reflecting the additional symmetry breaking other than the gauge symmetry $U$(1).
It is important to identify the pairing state with an internal degree of freedom of Cooper pairs
in order to examine exotic phenomena in unconventional superconductors (SCs).
Furthermore,
the identification gives the clue to the pairing mechanism
and offers the future application of unconventional SCs to new superconducting devices.


Recently, 
locally noncentrosymmetric (LNCS) superconducting systems are regarded as
a new family of exotic SCs \cite{yoshida2014kotai,doi:10.7566/JPSJ.83.061014}.
These systems are found in various real crystalline materials,
such as the heavy fermion superlattice CeCoIn$_5$/YbCoIn$_5$ \cite{mizukami2011},
the pnictide SC SrPtAs \cite{nishikubo2011},
the layered transition metal dichalcogenides \cite{PhysRevLett.108.196802} and so on.
A staggered anti-symmetric spin-orbit coupling (ASOC) due to the local noncentrosymmetricity appears
in the LNCS system with global inversion symmetry,
as the uniform ASOC appears in the NCS system without global inversion symmetry.
In LNCS systems, the sublattice structure plays an important role in the internal structure of the Cooper pair \cite{PhysRevB.84.184533}.
In multilayered system, 
Maruyama {\it et al.} showed that 
superconducting properties continuously change
from the isolated NCS layers to the correlated LNCS multilayer
with increasing the interlayer coupling,
in terms of the study of the spin susceptibility in the superconducting state \cite{maruyama2012}.
Exotic superconducting phases called the pair-density wave (PDW) phase \cite{yoshida2012}
and the complex-stripe phase \cite{yoshida2013}
were theoretically predicted to stabilize in multilayered systems in a high magnetic field,
when the paramagnetic depairing effect is dominant.
Indeed, experimental results in epitaxial superlattices CeCoIn$_{5}(n)$/YbCoIn$_{5}(5)$ suggested that
some exotic superconducting states might be realized at high magnetic fields \cite{PhysRevLett.109.157006}.
A numerical calculation demonstrated that
the PDW phase stabilizes at low temperatures and high magnetic fields
perpendicular to the layers through the $s$-wave pairing interaction,
layer-dependent Rashba ASOC and Zeeman magnetic field \cite{yoshida2012}.
In the PDW state, the phase of the superconducting order parameter modulates layer by layer \cite{yoshida2012}.
For instance, in the simplest model for multilayered systems, namely bilayer systems,
the order parameter $\varDelta$ changes its phase by $\pi$ between layers,
$(\varDelta_1,\varDelta_2)=(\varDelta,-\varDelta)$.
Such unusual stacking of order parameter is in sharp contrast to the conventional BCS state,
where the order parameter is uniform, $(\varDelta_1,\varDelta_2)=(\varDelta,\varDelta)$.
In real materials, vortices appear in the high magnetic field,
although the previous studies neglected them \cite{yoshida2012,yoshida2013}.
When the vortex density
is not large,
the phase modulation around vortices does not seriously affect the phase modulation
between the stacked layers.
Thus, the vortices induced by the orbital effect do not
play a crucial role in the stability of the PDW phase \cite{yoshida2012}. 
It is, however, important to investigate the vortex state
since the excitations around vortex cores dominate transport and thermodynamic properties.
%
%
Moreover, the observation of quasiparticle states around a vortex core gives us a variety of information
to identify the pairing state.
The quasiparticle states can be experimentally studied through the local quasiparticle density of states (LDOS),
which are obtained by scanning tunneling microscopy/spectroscopy (STM/STS) experiments
with high spatial and energy resolutions.

The first STM/STS measurement of the vortex core states focusing on the NCS superconductivity
was conducted on the $\beta$-pyrochlore osmate KOs$_2$O$_6$ \cite{PhysRevLett.101.057004}.
Recently,
the vortex bound states are experimentally explored also in NCS BiPd by STM/STS \cite{NatCommun.BiPd2015}.
A theoretically predicted spectroscopic feature of the parity-mixed superconducting state
due to the lack of an inversion center
is a two-gap structure of quasiparticle spectra in the bulk \cite{hayashi2006},
but any distinct spectroscopic evidence was not observed in these experiments.
In recent years, however,
vortex core states in materials with strong SOC are measured by STM/STS
in the context of topological superconductivity \cite{PhysRevLett.112.217001,PhysRevLett.114.017001,2014PhRvB..89j0501Y}.
Thus, it is an intriguing study to elucidate the effect of SOC on the electronic structure
around a quantum vortex in exotic SCs.


This paper is organized as follows.
First, in Sec. \ref{Sec:II}, we introduce multilayered superconducting systems.
In Sec. \ref{Sec:III}, we sketch the quasiclassical Green's function method
with use of the perturbative method for SOC.
Then, we demonstrate that the electronic structure around a vortex core in the PDW pairing state
is drastically different from that in the BCS state
in Sec. \ref{Sec:IV}.
We also elucidate
key roles of SOC on the difference in 
LDOS structure around a core between the BCS and PDW states.
In Sec. \ref{Sec:V},
we discuss energy spectra in the bulk and clarify
the relation to the vortex bound states.
In Sec. \ref{Sec:Discussions},
we discuss further the properties of the PDW state
and its realization in real crystalline materials.
A brief summary is given in Sec. \ref{Sec:VI}.

\section{Multilayered system\label{Multilayered system}}
\label{Sec:II}
In this paper,
we consider superconductivity in a bilayer system (the number of layers is $N=2$)
with a layer-dependent ASOC, $(\alpha_1,\alpha_2)=(\alpha,-\alpha)$,
the Zeeman field and the interlayer hopping.
Throughout the paper, we use the unit in which $\hbar=k_{\rm B}=c=1$.
The normal state is described by the following Hamiltonian including a spin degree of freedom:
\begin{align}
\mathscr{H}_0 = &\sum_{s,s^\prime,m} \int d\bm{r} \int d\bm{r}^\prime \psi^\dag_{sm}(\bm{r}) h_{ss^\prime m}\left( \bm{r},\bm{r}^\prime,-i\bm{\nabla}_{\bm{r}^\prime} \right) \psi_{s^\prime m}(\bm{r}^\prime)\nonumber\\
       &+t_\perp \sum_{s,\langle m,m^\prime \rangle} \int d\bm{r} \psi^\dag_{sm}(\bm{r}) \psi_{s m^\prime}(\bm{r}),\\
h_{ss^\prime m} &\left( \bm{r},\bm{r}^\prime,-i\bm{\nabla}_{\bm{r}^\prime} \right) = \delta_{ss^\prime} \delta(\bm{r}-\bm{r}^\prime)\xi(\bm{r}^\prime,-i\bm{\nabla}_{\bm{r}^\prime})\nonumber \\
&-\delta(\bm{r}-\bm{r}^\prime) \mu_{\rm B}\bm{H} \cdot \bm{\sigma}_{ss^\prime}+ \alpha_m \bm{g}(\bm{r}-\bm{r}^\prime) \cdot \bm{\sigma}_{ss^\prime},
\end{align}
where $\psi^\dag_{sm}(\bm{r})$ $[\psi_{sm}(\bm{r})]$
is the field operator creating (annihilating) a quasiparticle
with the spin $s$ at the position $\bm{r}$ in the $m$-th superconducting layer
in the Schr\"{o}dinger representation
and $\xi(\bm{r}^\prime,-i\bm{\nabla}_{\bm{r}^\prime})=\left[ -i\bm{\nabla}_{\bm{r}^\prime} +e\bm{A}(\bm{r}^\prime)\right]^2/(2m_{\rm e})-\mu$
is the free electron energy dispersion measured from the chemical potential $\mu$
with the bare electron mass $m_{\rm e}$,
the absolute value of the electron charge $e$
and the vector potential $\bm{A}(\bm{r})$.
$t_\perp$ is the interlayer coupling energy.
$\langle m,m^\prime \rangle$ indicates the summation over the neighboring layers.
$\mu_{\rm B}$ is the magnetic moment of quasiparticles,
and $\bm{H}=(0,0,H)$ is a magnetic field perpendicular to the superconducting layers.
$\bm{\sigma}=(\sigma_x,\sigma_y,\sigma_z)^{\rm T}$ is the vector representation of the Pauli spin matrix.
$\alpha_m$ is the spin-orbit coupling energy in the $m$-th layer and the orbital vector $\bm{g}(\bm{r}-\bm{r}^\prime)$ characterizing the SOC is defined through $\bm{g}(\bm{k})$ as
\begin{equation}
\bm{g}(\bm{k})=\int d\bar{\bm{r}}e^{-i\bm{k}\cdot \bar{\bm{r}}}\bm{g}(\bar{\bm{r}}=\bm{r}-\bm{r}^\prime).
\end{equation}
Here $\bm{k}$ denotes the relative wave vector.
We consider the Rashba type SOC in the two dimensional system described by the orbital vector
$\bm{g}(\bm{k})=(-k_y,k_x,0)/k_{\rm F}$ with the Fermi wave number $k_{\rm F}$ and
$(k_x,k_y)=k(\cos \phi_k,\sin \phi_k)$.
$\phi_k$ is the azimuthal angle.

In this study,
we neglect the mixing of Cooper pair wave functions with different parity for simplicity
and consider the spin-singlet $s$-wave pairing.
The superconducting order parameter
in $N$-layered systems
is expressed as $\hat{\varDelta}(\bm{r})=\varDelta(\bm{r})i\sigma_y \otimes D$,
where $D$ is the $N \times N$ diagonal matrix in the space composed of the layer degree of freedom (band space).
In the bilayer system ($N=2$), $D={\rm diag}(1,s)$
with $s=1~(-1)$ corresponding to the BCS (PDW) state.
A symbol $\hat{\cdot}$ denotes the $2N \times 2N$ matrix in spin and band space.
In this paper, we investigate vortex core structures
by assuming the layer dependence of order parameter (BCS or PDW state)
and leave the discussion of its thermodynamic stability to future studies.
Considering the SC with a dominant paramagnetic depairing effect and a large Gintzburg-Landau parameter,
we ignore the vector potential.
The vortex solution in this situation is studied by self-consistently determining 
the spatial profile of the pairing potential $\varDelta(\bm{r})$.


\section{Quasiclassical theory in multilayered systems}
\label{Sec:III}
In the PDW state,
the order parameter shows a spatial modulation
perpendicular to layers
in the length scale of the lattice constant,
which is much shorter than the characteristic length scale of most SCs.
On the other hand,
the order parameter varies in the scale of the coherence length within the layer.
Therefore, we can develop the quasiclassical theory as follows.

We investigate the electronic structure around a single vortex
by means of the quasiclassical theory.
As a result of the quasiclassical approximation,
the quasiclassical Green's function depends on the center of mass coordinate of the Cooper pair $\bm{r}$,
the direction of the relative wave vector (or momentum)
of the Cooper pair $\tilde{\bm{k}}=(\cos \phi_k,\sin \phi_k)$,
and the Matsubara frequency for fermions $\omega_n=(2n+1)\pi T$ with the temperature $T$ and an integer $n$.
We define the quasiclassical Green's function as the following $4N \times 4N$ matrix in the Nambu space:
\begin{equation}
\check{g}(\bm{r},\tilde{\bm{k}},i\omega_n)=-i\pi
\left(
\begin{array}{cc}
\hat{g}(\bm{r},\tilde{\bm{k}},i\omega_n) & i\hat{f}(\bm{r},\tilde{\bm{k}},i\omega_n) \\
-i\hat{\bar{f}}(\bm{r},\tilde{\bm{k}},i\omega_n) & -\hat{\bar{g}}(\bm{r},\tilde{\bm{k}},i\omega_n) 
\end{array}
\right).
\end{equation}

Let us derive the Eilenberger equation with the $4 N \times 4 N$ matrix quasiclassical Green's function
using the perturbation method \cite{nagai2015}.
In the absence of the Rashba ASOC, Zeeman magnetic field and interlayer hopping,
({\it i.e.,} $\alpha=\mu_{\rm B}H=t_{\perp}=0$),
the FSs have $2N$-fold degeneracy in the normal state. 
In this case, one can easily obtain the unperturbed Eilenberger equation expressed as
\begin{align}
i\bm{v}_{\rm F}(\tilde{\bm{k}})& \cdot \bm{\nabla}\check{g}(\bm{r},\tilde{\bm{k}},i\omega_n)
\nonumber \\
&+
\Bigl[
i\omega_n \check{\tau}_3
-\check{\varDelta}(\bm{r})
,~\check{g}(\bm{r},\tilde{\bm{k}},i\omega_n)
\Bigr]
=\check{0},
\end{align}
where
\begin{align}
\check{\tau}_3
 &=
 \left(
\begin{array}{cc}
\sigma_0 \otimes I_{N \times N} & \hat{0} \\
\hat{0} & -\sigma_0 \otimes I_{N \times N}
\end{array}
\right),
\\
\check{\varDelta}(\bm{r})
 &= 
\left(
\begin{array}{cc}
\hat{0} & \hat{\varDelta}(\bm{r}) \\
-\hat{\varDelta}^\dag(\bm{r}) & \hat{0}
\end{array}
\right).
\end{align}
Here,
$\sigma_0$ and $I_{N \times N}$ are the unit matrices in spin and band spaces, respectively.
The braket $[\cdots,\cdots]$ is a commutator.
We add the Zeeman, Rashba, and interlayer hopping terms into the above equations through a self-energy as 
\begin{align}
\check{K}(\tilde{\bm{k}})
&= \left[ -\mu_{\rm B} \bm{H}
   +\alpha \check{\bm{g}}(\tilde{\bm{k}}) \right] \cdot \check{\bm{S}}
   +t_\perp(\sigma_0 \otimes T_\perp) \otimes \check{\tau}_0,\\
\check{\bm{S}}
 &=
\left(
\begin{array}{cc}
\bm{\sigma}\otimes I_{N \times N} & \hat{0} \\
\hat{0} & \bm{\sigma}^\ast \otimes I_{N \times N}
\end{array}
\right),
\\
\check{\bm{g}}(\tilde{\bm{k}})
 &=
\left(
\begin{array}{cc}
\bm{g}(\tilde{\bm{k}})\sigma_0 \otimes S_{\rm d} & \hat{0}  \\
\hat{0} & \bm{g}(-\tilde{\bm{k}})\sigma_0 \otimes S_{\rm d}
\end{array}
\right).
\end{align}
Here, $T_\perp={\rm offdiag}(1,1)$ and $S_{\rm d}={\rm diag}(1,-1)$. 
offdiag$(\cdot,\cdot)$ denotes the $2 \times 2$ matrix which has only the offdiagonal component in the band space.
Thus, we obtain the Eilenberger equation in the $4N \times 4N$ matrix form with the self-energy matrix $\check{K}(\tilde{\bm{k}})$ as
\begin{align}
i\bm{v}_{\rm F}&(\tilde{\bm{k}}) \cdot \bm{\nabla}\check{g}(\bm{r},\tilde{\bm{k}},i\omega_n)
\nonumber \\
&+
\Bigl[
i\omega_n \check{\tau}_3
-\check{\varDelta}(\bm{r})
-\check{K}(\tilde{\bm{k}})
,~\check{g}(\bm{r},\tilde{\bm{k}},i\omega_n)
\Bigr]
=\check{0}.
\label{Eilenberger eq.}
\end{align}

Since we consider the SC in which the paramagnetic depairing effect is dominant,
we incorporate the Zeeman term into the Eilenberger equation.
In the presence of the Zeeman term,
the Eilenberger equation cannot be decomposed into the two decoupled equations
for the spin-split Fermi surface
although we can do so at $H=0$ by using the band basis representation \cite{hayashi-SFD-2006}.
Therefore, we take the orbital basis,
in which the spin quantization axis is parallel to the $z$ axis.
In the orbital basis,
we can transform Eq.~(\ref{Eilenberger eq.}) into the two matrix Riccati equations
regarding $\check{K}(\tilde{\bm{k}})$ as the {\it self energy} \cite{higashi2014}:
\begin{align}
\bm{v}_{\rm F}(\tilde{\bm{k}}) \cdot \bm{\nabla}\hat{a}_0
 + 2\omega_n \hat{a}_0&+\hat{a}_0 \hat{\varDelta}^\dag_0 \hat{a}_0 -\hat{\varDelta}_0 \nonumber \\
&+i\left( \hat{K}^0_{11}\hat{a}_0+\hat{a}_0\hat{K}^0_{22} \right) = \hat{0},
\\
\bm{v}_{\rm F}(\tilde{\bm{k}}) \cdot \bm{\nabla}\hat{b}_0
 - 2\omega_n \hat{b}_0&-\hat{b}_0\hat{\varDelta}_0\hat{b}_0 + \hat{\varDelta}^\dag_0 \nonumber \\
 &-i \left( \hat{b}_0\hat{K}^0_{11}+\hat{K}^0_{22}\hat{b}_0  \right) = \hat{0}.
\end{align}
Here we define $\check{K}(\tilde{\bm{k}})={\rm diag}(\hat{K}_{11}(\tilde{\bm{k}}),\hat{K}_{22}(\tilde{\bm{k}}))$ in the Nambu space
with $\hat{K}_{11}(\tilde{\bm{k}}) \equiv -\mu_{\rm B}H \sigma_z \otimes I_{N \times N} + \alpha \bm{g}(\tilde{\bm{k}}) \cdot \bm{\sigma} \otimes S_{\rm d} +t_\perp \sigma_0 \otimes T_\perp$
and  $\hat{K}_{22}(\tilde{\bm{k}}) \equiv -\mu_{\rm B}H \sigma_z \otimes I_{N \times N} - \alpha \bm{g}(\tilde{\bm{k}}) \cdot \bm{\sigma}^\ast \otimes S_{\rm d} +t_\perp \sigma_0 \otimes T_\perp$,
and we introduce the following expressions,
$\hat{a}=\hat{a}_0( i\sigma_y \otimes I_{N \times N} )$,
$\hat{b}=(i\sigma_y \otimes I_{N \times N}) \hat{b}_0$,
$\hat{K}_{11}=\hat{K}^0_{11}$,
$\hat{K}_{22}=(-\sigma_y \otimes I_{N \times N})\hat{K}^0_{22}(\sigma_y \otimes I_{N \times N})$,
$\hat{\varDelta}=\hat{\varDelta}_0(i\sigma_y \otimes I_{N \times N})$
and
$\hat{\varDelta}^\dag=(i\sigma_y \otimes I_{N \times N})\hat{\varDelta}^\dag_0$.
$\hat{a}$ and $\hat{b}$ are the former Riccati parameters.

Using the pairing interaction adopted by Ref.~\cite{frigeri2006},
the gap equation for the spin-singlet component is obtained as \cite{RevModPhys.63.239}
\begin{align}
\varDelta(\bm{r})=\lambda \pi T \cfrac{1}{2}
&
\sum_{-n_{\rm c}(T)-1 < n < n_{\rm c}(T)}
\sum_{s^\prime_1 s^\prime_2}(i \sigma_y)^\dag_{s^\prime_2 s^\prime_1}
\nonumber \\
&\times \left\langle f^0_{s^\prime_1 s^\prime_2}(\bm{r},\tilde{\bm{k}}^\prime,i\omega_n) (i \sigma_y)_{s^\prime_1 s^\prime_2} \right\rangle_{\tilde{\bm{k}}^\prime},
\end{align}
where $\langle \cdots \rangle_{\tilde{\bm{k}}}$ is the average on the Fermi surface .
We use the following coupling constant obtained in the bulk at $\alpha=\mu_{\rm B}H=t_\perp=0$:
\begin{equation}
\cfrac{1}{\lambda}=\ln \left(\cfrac{T}{T_{\rm c0}}\right) + \sum_{0 \leq  n < n_{\rm c}(T)}\cfrac{2}{2n+1},
\end{equation}
where $T_{\rm c0}$ is the superconducting transition temperature
at $\alpha=\mu_{\rm B}H=t_\perp=0$ and $n_{\rm c}(T)=(\omega_{\rm c}/\pi T-1)/2$.
We fix the cutoff frequency to $\omega_{\rm c}=7\pi T_{\rm c0}$.
The LDOS per spin and layer is given by
\begin{equation}
N(E,\bm{r})=-\cfrac{N_{\rm F}}{2N} \cfrac{1}{\pi}\left\langle {\rm Im} \left[{\rm Tr}\hat{g}(\bm{r},\tilde{\bm{k}},i\omega_n \rightarrow E+i\eta) \right] \right\rangle_{\tilde{\bm{k}}}.
\end{equation}
$N_{\rm F}$ is the DOS per spin and layer at the Fermi level in the normal state.
$E$ and $\eta$ are the real energy and the energy smearing factor, respectively.

\section{Electronic structure around a vortex core}
\label{Sec:IV}
In this section, 
we clarify
the difference of the electronic structure around a vortex core between the BCS and PDW states.
We show the self-consistently calculated spatial profiles of the pair potential and LDOS.
The presence or absence of zero energy peak (ZEP) in LDOS is demonstrated.
We fix the temperature, the Zeeman field and the interlayer hopping to
$T/T_{{\rm c}0}=0.1$, $\mu_{\rm B}H/T_{{\rm c}0}=1.5$ and $t_\perp/T_{{\rm c}0}=1$, respectively,
unless we show explicitly.

\subsection{Pair potential}
\begin{figure}[tb]
  \begin{center}
    \begin{tabular}{p{85mm}}
      \resizebox{85mm}{!}{\includegraphics{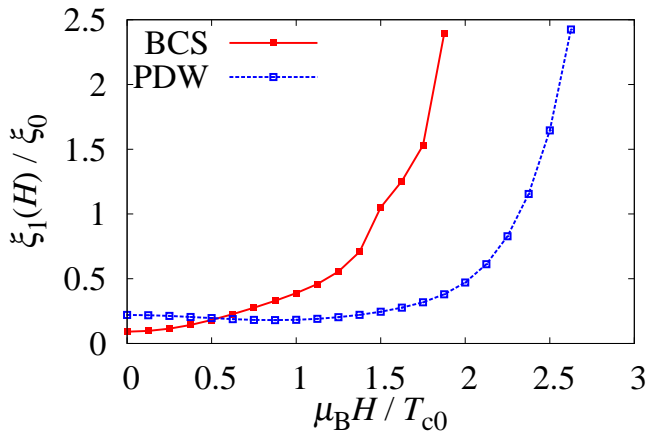}}
    \end{tabular}
\caption{
\label{fig1}
(Color online)
Zeeman magnetic field dependences of the vortex core radius $\xi_1(H)$ for $\alpha/T_{\rm c0}=2$.
We set $T/T_{\rm c0}=0.1$ and $t_\perp/T_{\rm c0}=1$.
}
  \end{center}
\end{figure}

\begin{figure}[tb]
  \begin{center}
    \begin{tabular}{p{85mm}p{85mm}}
      \resizebox{85mm}{!}{\includegraphics{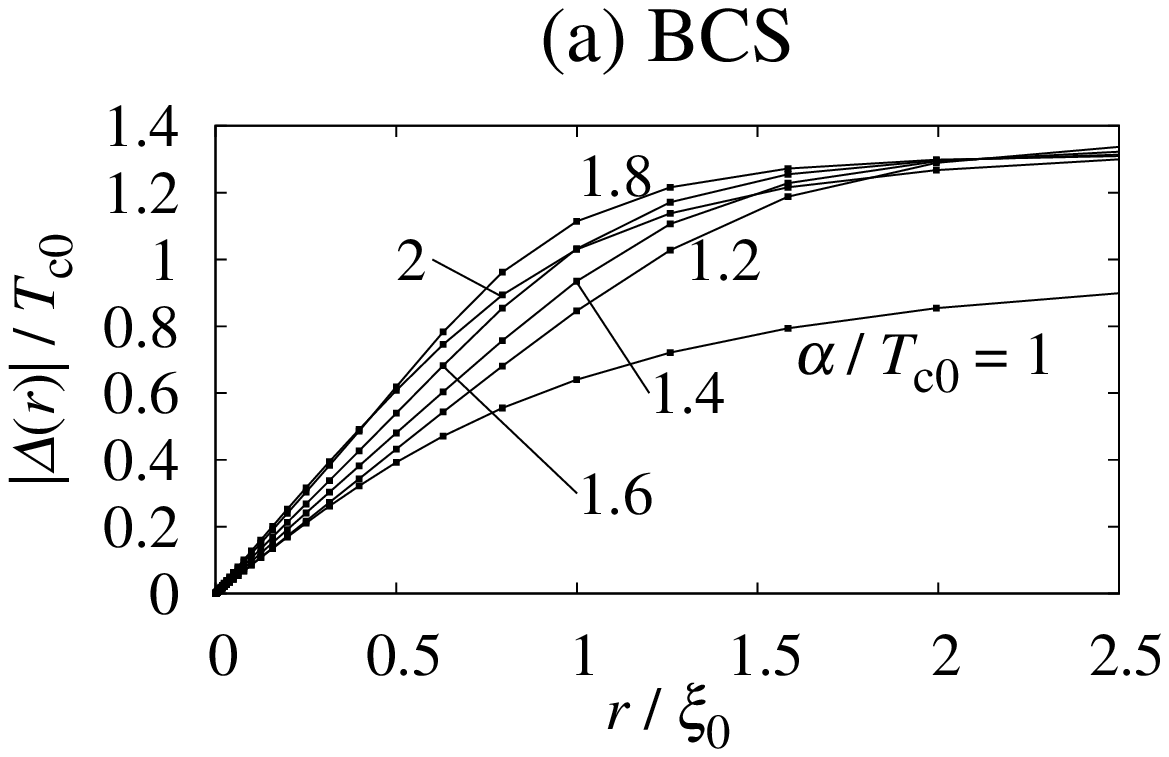}} \\
      \resizebox{85mm}{!}{\includegraphics{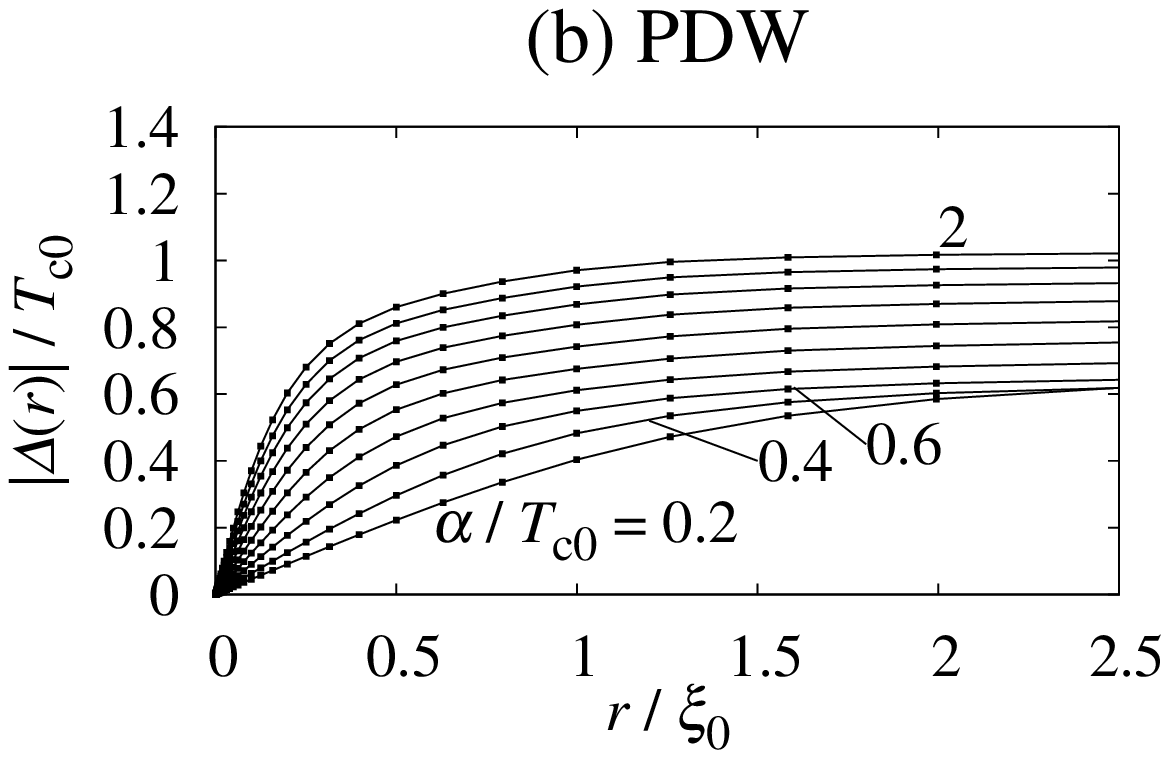}} 
    \end{tabular}
\caption{
\label{fig2}
Spatial profiles of the pair potential amplitudes $|\varDelta(r)|$ for the (a) BCS and (b) PDW states.
The horizontal axis represents the radial distance from a vortex center $r=0$.
We set $T/T_{\rm c0}=0.1$ and $t_\perp/T_{\rm c0}=1$.
SOC strength is indicated for each curve.
$\alpha/T_{{\rm c}0}=0.2-2$ is increased from the bottom to the top by 0.2 step in (b).
}
  \end{center}
\end{figure}

Let us discuss the spatial profiles of the pair potential amplitude $|\varDelta(r)|$ around a vortex.

First, we discuss the Zeeman magnetic field dependence of the vortex core radius for $\alpha/T_{\rm c0}=2$.
The vortex core radius is defined as \cite{kramer1974,sonier,hayashi2005}
\begin{equation}
\xi_1(H)=\varDelta(H,r=r_{\rm c})/\lim_{r \to 0}\cfrac{\varDelta(H,r)}{r},
\end{equation}
where we set $r_{\rm c}=10\xi_0$.
As shown in Fig.~\ref{fig1},
the vortex core radius $\xi_1(H)$ in the BCS state diverges at the critical magnetic field
due to the paramagnetic depairing.
Because the PDW state is more robust against the paramagnetic depairing than the BCS state,
the critical magnetic field of the PDW state is higher than that of the BCS state.
Indeed, Fig.~\ref{fig1} shows the divergence of $\xi_1(H)$ in the PDW state
at a higher magnetic field.
Then, the vortex core suddenly shrinks at the first-order BCS-PDW transition
which occurs at
$\mu_{\rm B}H/T_{\rm c0} \simeq 1.8$ \cite{higashi2014multilayer}.
This sudden shrinkage of vortex core
originates from the difference of the superconducting gap
between the BCS and PDW states.
At the transition magnetic field,
the superconducting gap is larger in the PDW state than in the BCS state \cite{yoshida2012}.

Next, we discuss the effect of SOC on the spatial profiles of pair potential.
The pair potential amplitude $|\varDelta(r)|$
significantly
depends on the strength of the SOC, as shown in Fig.~\ref{fig2}.
The horizontal axis indicates the radial distance from the vortex center ($r=0$)
normalized by the coherence length $\xi_0=v_{\rm F}/T_{{\rm c}0}$
for $\alpha=\mu_{\rm B}H=t_{\perp}=0$.
The BCS state [Fig.~\ref{fig2}(a)] is destabilized by the paramagnetic depairing
for a small SOC $\alpha/T_{{\rm c}0} \lesssim 1$
because we adopt a magnetic field larger than the conventional Pauli limit.
On the other hand,
the PDW state is robust against the magnetic field,
since the paramagnetic depairing is suppressed \cite{maruyama2012} and
the PDW state survives even when the SOC is small $\alpha/T_{{\rm c}0} \lesssim 1$ [Fig.~\ref{fig2}(b)].
It is shown that both BCS and PDW states are stabilized
by the SOC in the high magnetic field ($\mu_{\rm B}H/T_{{\rm c}0}=1.5$).
In the PDW state,
the pair potential amplitude gets larger monotonically with increasing the SOC strength $\alpha$.
This is because the PDW state can be mapped onto the two-dimensional Rashba SC
with use of the mirror symmetry as discussed later,
and then the Rashba type SOC locks the spin quantization axis within the $x-y$ plane
to suppress the paramagnetic depairing effect.
In contrast, in the BCS state, 
the pair potential amplitude
shows an unusual non-monotonic behavior with increasing the SOC strength.

An intriguing feature is seen in the vortex core radius.
As shown in Fig.~\ref{fig2}(b),
the core radius in the PDW state is smaller than that in the BCS state
in accordance with Fig.~\ref{fig1}.
This feature gets prominent with increasing the SOC strength.



\subsection{Local density of states}
\begin{figure}[tb]
  \begin{center}
    \begin{tabular}{p{85mm}p{85mm}}
      \resizebox{85mm}{!}{\includegraphics{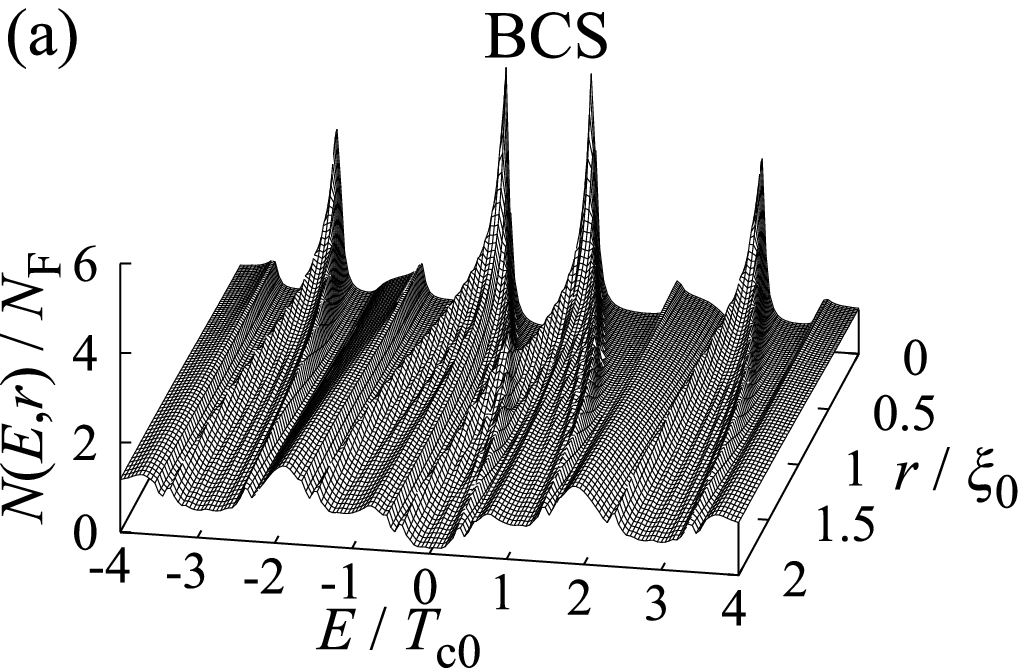}} \\
      \resizebox{85mm}{!}{\includegraphics{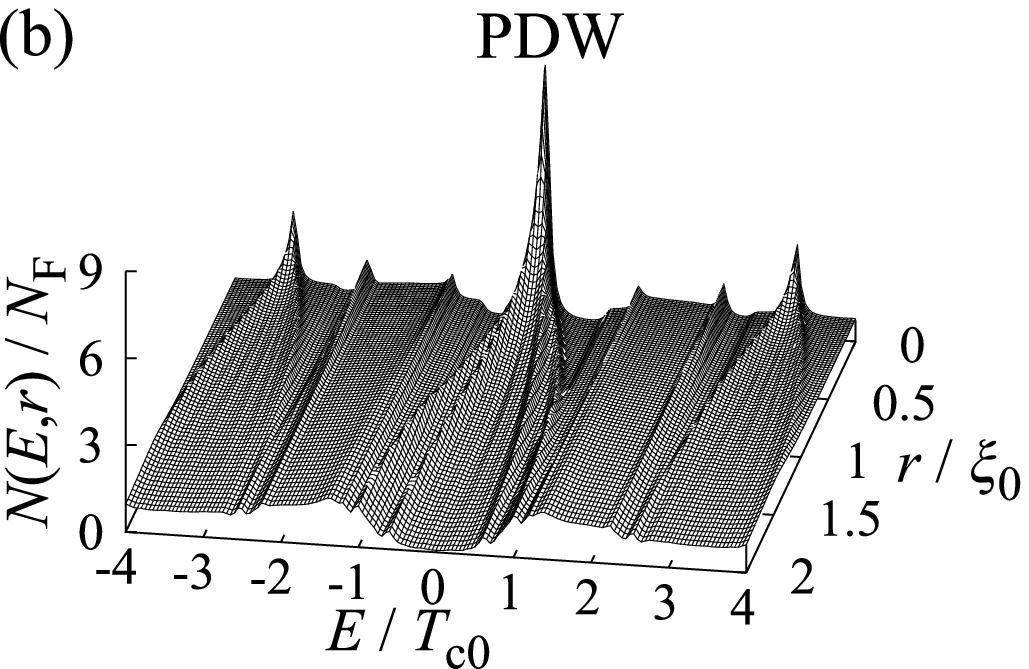}} 
    \end{tabular}
\caption{
\label{fig3}
LDOS $N(E,r)$
in the BCS state (a) and in the PDW state (b).
The Zeeman field is set to $\mu_{\rm B}H/T_{\rm c0}=1.5$ and the SOC is $\alpha/T_{\rm c0}=2$ for both pairing states.
Other parameters are $T/T_{\rm c0}=0.1$, $t_\perp/T_{\rm c0}=1$ and $\eta=0.05T_{\rm c0}$.
}
  \end{center}
\end{figure}
We here discuss the energy and spatial dependence of the LDOS around a vortex in the BCS and PDW states
and clarify the change of the electronic structure at the BCS-PDW transition. 
We fix the Zeeman magnetic field and the SOC strength to
$\mu_{\rm B}H/T_{\rm c0}=1.5$ and $\alpha/T_{{\rm c}0}=2$, respectively. 
The self-consistent solutions for the gap equation displayed in Fig.~\ref{fig2} are used to calculate the LDOS.
In the BCS states, as shown in Fig.~\ref{fig3}(a),
the zero energy vortex bound states split into the four peaks due to the interlayer hopping and the Zeeman field.
On the other hand, in the PDW state,
a large quasiparticle DOS appears at the zero energy [see Fig.~\ref{fig3}(b)].
This is quite contrasting with the LDOS structure in the BCS state.

When we consider
the PDW state in the absence of
the SOC (The PDW state is indeed unstable in the absence of the SOC),
the four LDOS peaks appear
as in the BCS state due to the interlayer hopping and the Zeeman field.
The magnetic field dependence of the LDOS is also similar to that in the BCS state. 

Thus, the contrasting behaviors between the PDW and BCS states in Fig.~\ref{fig3} results from the SOC.
One might speculate
that
the two peaks in the LDOS at $\alpha=0$ get combined with increasing the SOC strength.
However, we should notice
another origin of the appearance of the zero energy vortex bound states in the PDW state,
which is described in the remaining part of this section.

\subsection{Emergent zero energy peak by spin-orbit coupling}
\begin{figure}[tb]
  \begin{center}
    \begin{tabular}{p{85mm}p{85mm}}
      \resizebox{85mm}{!}{\includegraphics{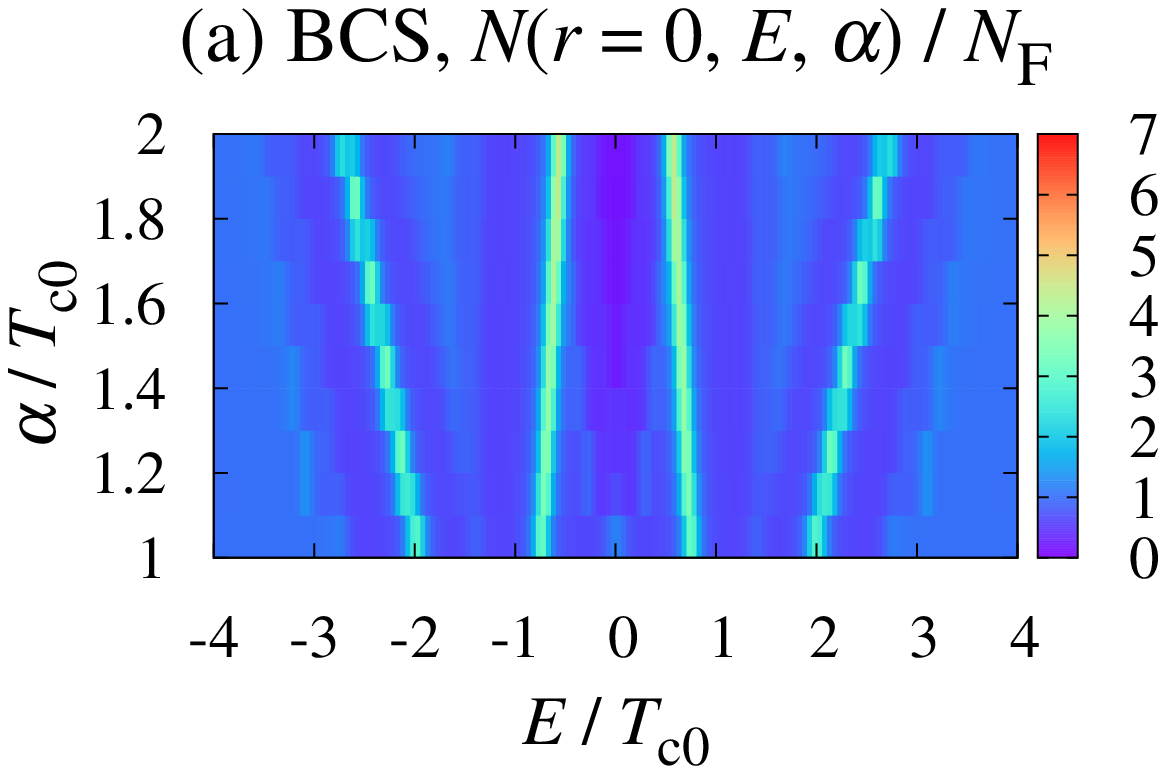}} \\
      \resizebox{85mm}{!}{\includegraphics{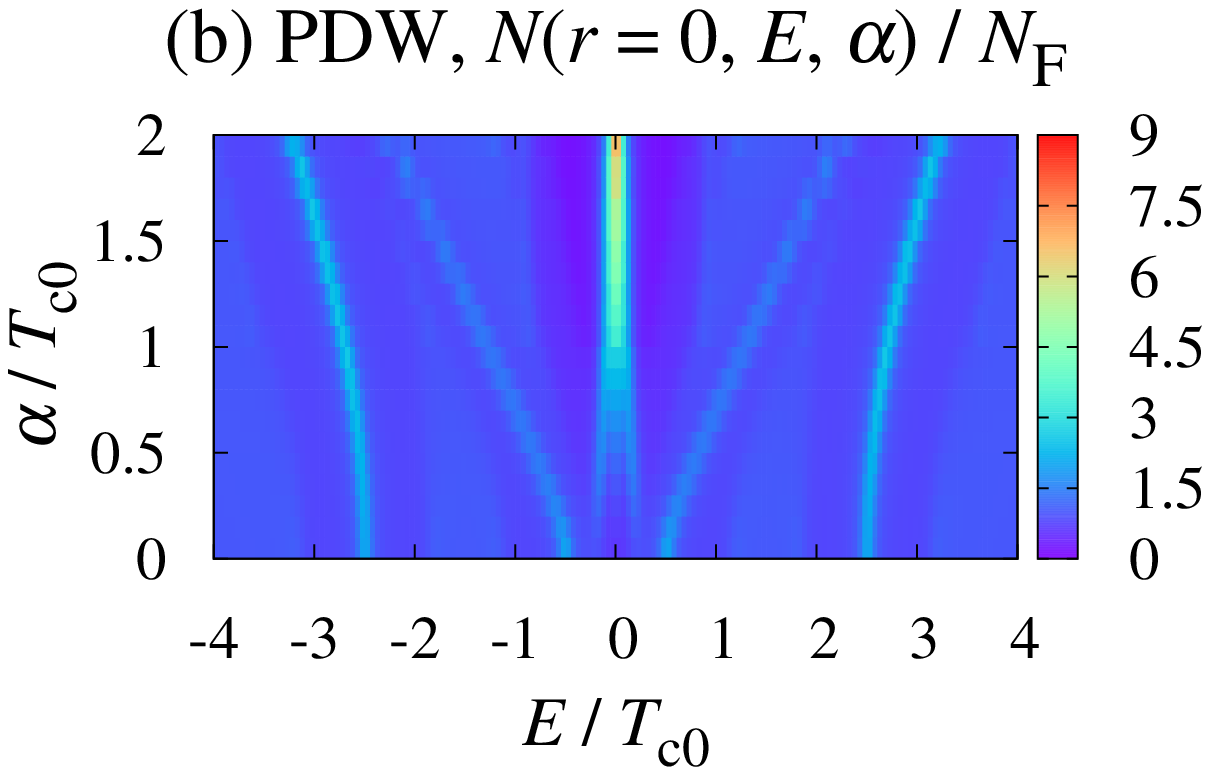}}
    \end{tabular}
\caption{
\label{fig4}
(Color online)
The energy and SOC strength dependences of the LDOS
at the vortex center $N(r=0,E,\alpha)$ in the BCS state (a) and in the PDW state (b)
for $\mu_{\rm B}H/T_{\rm c0}=1.5$.
Other parameters are $T/T_{\rm c0}=0.1$, $t_\perp/T_{\rm c0}=1$ and $\eta=0.05T_{\rm c0}$.
}
  \end{center}
\end{figure}
In this subsection,
we show the SOC strength dependence of the vortex bound states.
As we discussed in the previous subsection,
in the absence of the SOC,
the quasiparticle structure around the vortex core is similar between the PDW and BCS states.
On the other hand,
in the presence of the SOC,
the internal structure of Cooper pairs manifests itself in the quasiparticle structure \cite{PhysRevB.84.184533,maruyama2012,yoshida2012}.
Indeed, the zero energy quasiparticle state appears
around a vortex core in the PDW state
as a result of the sign change of order parameter between layers [Fig.~\ref{fig3}(b)],
although the ZEP splits in the BCS state.
In order to elucidate the effect of the SOC on the quasiparticle structure,
we show in Fig.~\ref{fig4} the LDOS at the vortex center for various SOC strength.

For the parameters in Fig.~\ref{fig4},
the BCS state is completely suppressed due to the paramagnetic depairing
for $\alpha/T_{{\rm c}0} \lesssim 1$.
Thus, we show the results for $\alpha/T_{\rm c0}\geq 1$.
As shown in Fig.~\ref{fig4}(a),
Andreev bound states have a finite energy almost independent of the SOC strength.
The peaks in the low energy region shift a little to lower energy with increasing the SOC strength,
whereas those in the high energy region move to higher energy.

On the other hand, in the PDW state,
the ZEP in LDOS gradually
develops with increasing the SOC strength [see Fig.~\ref{fig4}(b)].
At $\alpha/T_{\rm c0}=2$, the LDOS
at the vortex center
clearly shows the ZEP, which has already been shown in Fig.~\ref{fig3}(b).
Thus, the SOC plays a crucial role in the contrasting quasiparticle structure
around a vortex core between the BCS and PDW states.
The SOC is much larger than the superconducting gap in most NCS and LNCS,
and thus, the condition, $\alpha/T_{\rm c0} \gg 1$, is satisfied.
Therefore, the zero energy bound states will appear around the vortex core
when the PDW state is stabilized in the magnetic field.

Although the vortex core states in the BCS state show Zeeman splitting,
the ZEP in the PDW state is robust against the Zeeman field.
This result can be viewed as a result of the suppression of the paramagnetic depairing effect in the PDW state.
An indication for the suppression of the paramagnetic depairing effect is
obtained by calculating the $c$-axis spin susceptibility in the superconducting state.
As shown in Ref.~\cite{maruyama2012}, the spin susceptibility is not decreased by the PDW order
(see Figs. 2 and 11 in Ref.~\cite{maruyama2012}).
This indicates that the Zeeman field does not suppress the superconducting state
and does not split the ZEP.
Another consequence of the suppression of the paramagnetic depairing effect
is the particle-hole symmetry in the mirror subsector in the PDW state \cite{PhysRevLett.115.027001}.
This aspect is discussed in Sec. \ref{Sec:V}.

In the next section,
we show the energy spectra in the bulk to clarify the effect of the SOC on the energy dispersion.
We also discuss the relation between the vortex bound states and the bulk energy spectra.

\section{Energy spectra in the bulk}
\label{Sec:V}
In this section,
we investigate effects
of the SOC on the energy spectra in the bulk
and discuss the relation between the bulk superconducting gap and the vortex bound states.
We diagonalize the following $8\times8$ Bogoliubov-de Gennes (BdG) Hamiltonian to obtain the energy spectra:
\begin{align}
\check{\mathscr{H}}_{\rm BdG}
&=
\left(
\begin{array}{cc}
\hat{\mathscr{H}}_0(\bm{k}) & \hat{\varDelta}  \\
\hat{\varDelta}^\dag& -\hat{\mathscr{H}}^\ast_0 (\bm{-k})
\end{array}
\right),\\
\hat{\mathscr{H}}_0(\bm{k})
&=
\left(
\begin{array}{cc}
h_1(\bm{k}) & t_\perp \sigma_0  \\
t_\perp \sigma_0 & h_2(\bm{k})
\end{array}
\right),
\end{align}
where $h_m(\bm{k})=\xi(k)\sigma_0+\alpha_m \bm{g}(\bm{k}) \cdot \bm{\sigma}-\mu_{\rm B}\bm{H} \cdot \bm{\sigma}$.
We adopt the isotropic dispersion relation $\xi(k)=k^2/2m_{\rm e} -\mu$ and
the orbital vector
$\bm{g}(\bm{k})=(-k_y,k_x,0)/k_{\rm F0}=(k/k_{\rm F0})(-\sin \phi_k,\cos \phi_k,0)$ in two dimensional layers.
$k_{\rm F0}$ is the Fermi wave number for $\alpha=\mu_{\rm B}H=t_\perp=0$.

In most NCS and LNCS,
the SOC strength $\alpha$ is much larger than the superconducting gap energy at zero temperature $\varDelta_0$
and much smaller than the Fermi energy $E_{\rm F}$.
Thus, the SCs are in the quasiclassical regime, namely, $k_{\rm F}\xi_0\sim E_{\rm F}/\varDelta_0\gg1$,
and the energy scale of the SOC may satisfy the condition $\varDelta_0 \ll \alpha \ll E_{\rm F}$.
However, we adopt parameters in the quantum limit regime ($E_{\rm F}/\varDelta_0=5$)
for the visibility of figures.
We confirmed that the following discussions are correct also in the quasiclassical regime.
\subsection{Absence of spin-orbit coupling}
\begin{figure}[tb]
  \begin{center}
    \begin{tabular}{p{85mm}p{85mm}}
      \resizebox{85mm}{!}{\includegraphics{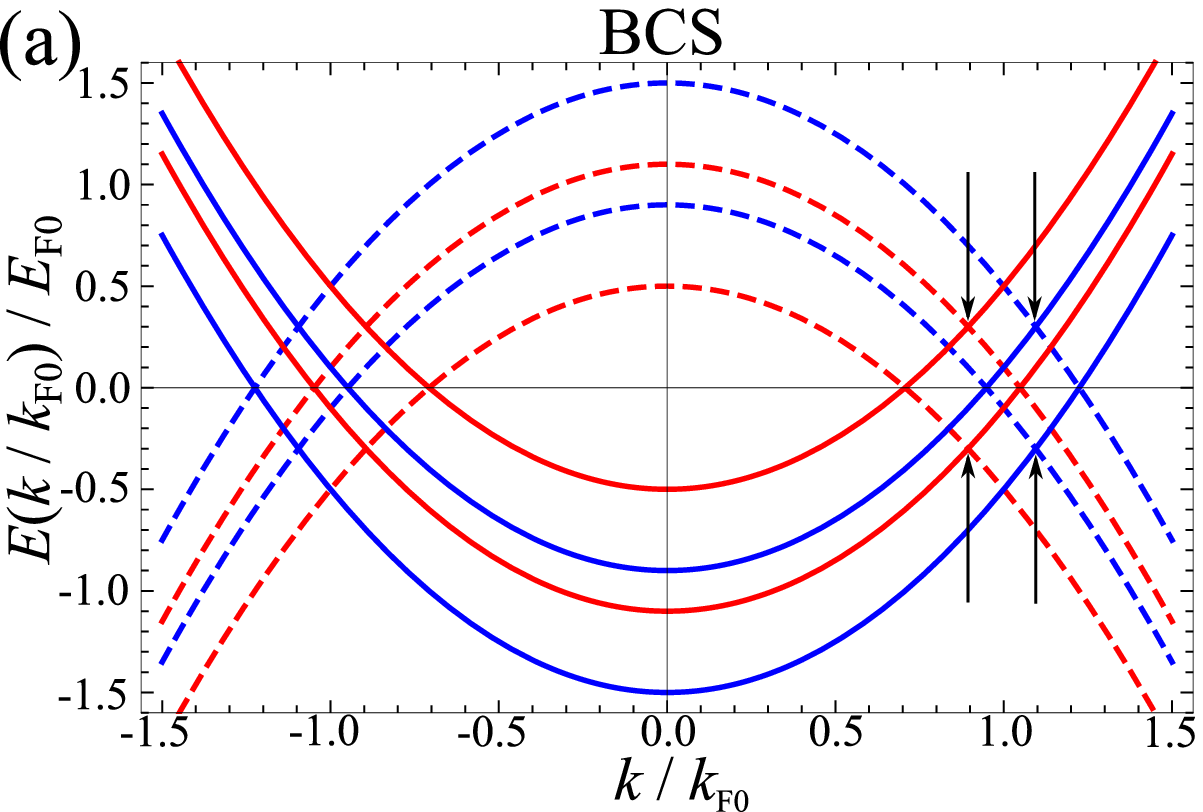}}\\
      \resizebox{85mm}{!}{\includegraphics{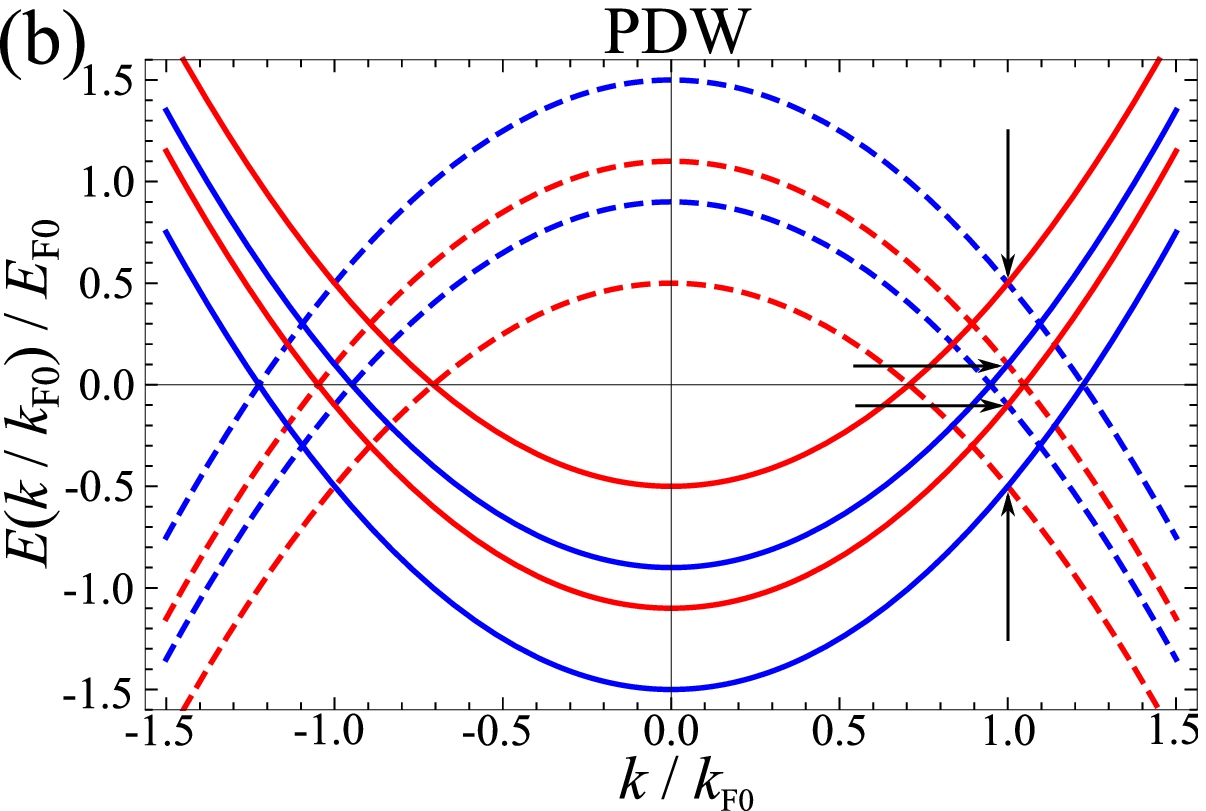}}
    \end{tabular}
\caption{
\label{fig5} (Color online)
Normal state energy spectra at $\alpha=0$.
Arrows indicate the intersections of electron and hole bands
which form Cooper pairs
in the BCS state (a) and the PDW state (b).
The horizontal axis denotes the wave number normalized by $k_{{\rm F}0}$.
Other parameters are set to $\mu_{\rm B}H/E_{\rm F0}=0.3$, $t_\perp/E_{\rm F0}=0.2$ and $\mu/E_{\rm F0}=1$
with $E_{\rm F0}=k^2_{\rm F0}/2m_{\rm e}$.
} 
  \end{center}
\end{figure}
The superconducting energy gap can be viewed
as a hybridization gap between electron and hole bands in a certain basis.
We study the $s$-wave BCS state and PDW state in this section.
As carried out in Refs.~\cite{ueno2013} and \cite{PhysRevLett.115.027001},
the BdG Hamiltonian is block-diagonalized by using the mirror symmetry.
Then, the
interlayer hopping $t_\perp$ is taken into account through
an effective Zeeman magnetic field
$h_\pm\sigma_z=(\mu_{\rm B}H \pm t_\perp)\sigma_z$ \cite{PhysRevLett.115.027001}
in a subsector Hamiltonian.
As a result of the lifting of four fold degenecary due to the effective Zeeman field,
four energy bands appear in
both electron and hole branches.
Figures \ref{fig5}(a) and \ref{fig5}(b) show
the normal state energy bands
$\pm E_{2\uparrow}(k)=\pm[\xi(k)-h_+]$,
$\pm E_{1\uparrow}(k)=\pm[\xi(k)-h_-]$,
$\pm E_{2\downarrow}(k)=\pm[\xi(k)+h_-]$ and
$\pm E_{1 \downarrow}(k)=\pm[\xi(k)+h_+]$
from bottom (top) to top (bottom) at $k=0$ for the electron (hole) bands.
The blue (gray) and red (light gray) lines show the energy bands for $m=2$ and 1, respectively.
We choose the parameters $\mu_{\rm B}H/T_{\rm c0}=1.5$ and $t_\perp/T_{\rm c0}=1$
($\mu_{\rm B}H/E_{\rm F0}=0.3$ and $t_\perp/E_{\rm F0}=0.2$).

As investigated in Ref.~\cite{maruyama2012} and pointed out in Ref.~\cite{yoshida2012},
in the absence of the SOC,
intraband quasiparticle states form the Cooper pairs in the BCS state,
whereas interband pairing states are realized in the PDW state.
In Figs.~\ref{fig5}(a) and \ref{fig5}(b),
arrows show the positions of superconducting gap in the BCS and PDW states, respectively.
In the BCS state
four spectral gaps open by
the intra band spin-singlet Cooper pairing [Fig.~\ref{fig5}(a)].
In the PDW state
four gaps are induced by
the inter band spin-singlet pairing [Fig.~\ref{fig5}(b)].
The superconducting gaps are symmetric
with respect to $E_{\rm F}$,
because of the particle-hole symmetry.
In both BCS and PDW states
the superconducting gaps
are shifted away from $E_{\rm F}$.
In the BCS state
at $\alpha=0$
the shift is due to the paramagnetic depairing effect,
and indeed, the BCS state
is completely destroyed due to the paramagnetic depairing at $\mu_{\rm B}H/T_{\rm c0}=1.5$.
On the other hand,
the PDW state
is unstable at $\alpha=0$
because the inter band pairing gives rise to the superconducting gap away from the Fermi energy even at $H=0$.


\subsection{Presence of spin-orbit coupling}
\begin{figure}[tb]
  \begin{center}
    \begin{tabular}{p{85mm}p{85mm}}
      \resizebox{85mm}{!}{\includegraphics{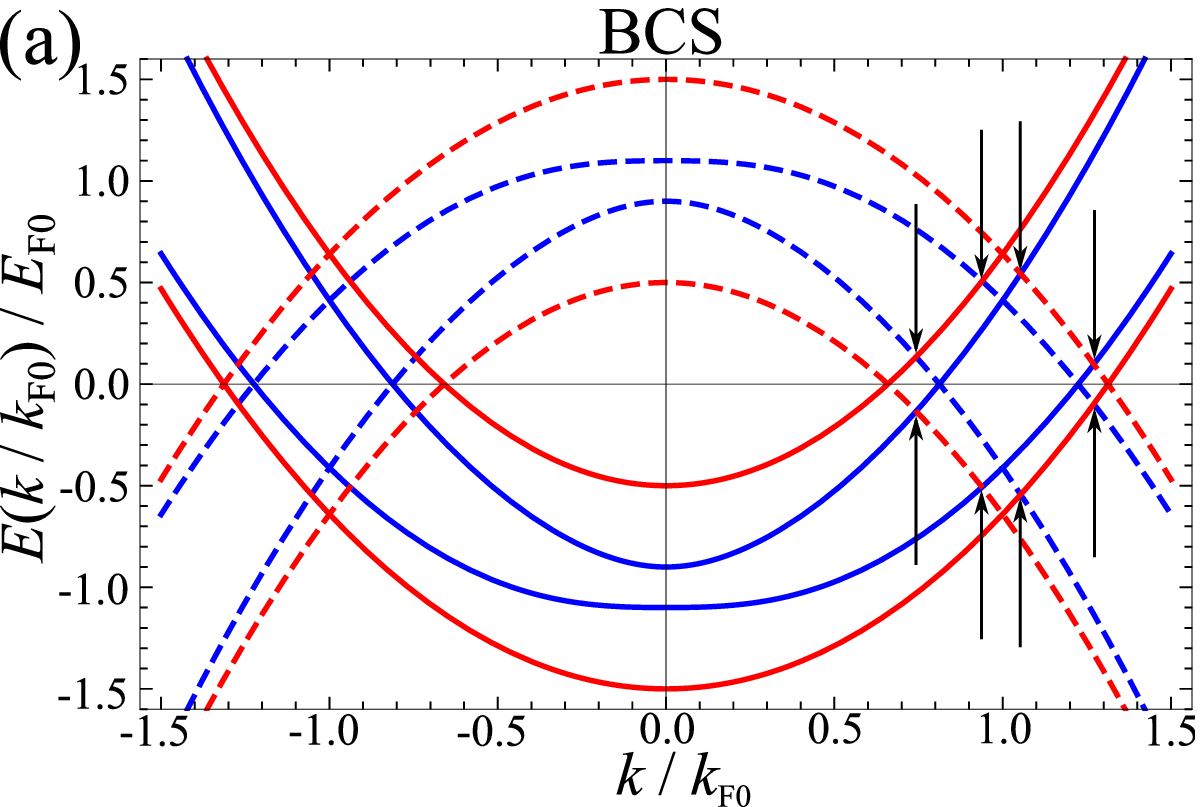}}\\
      \resizebox{85mm}{!}{\includegraphics{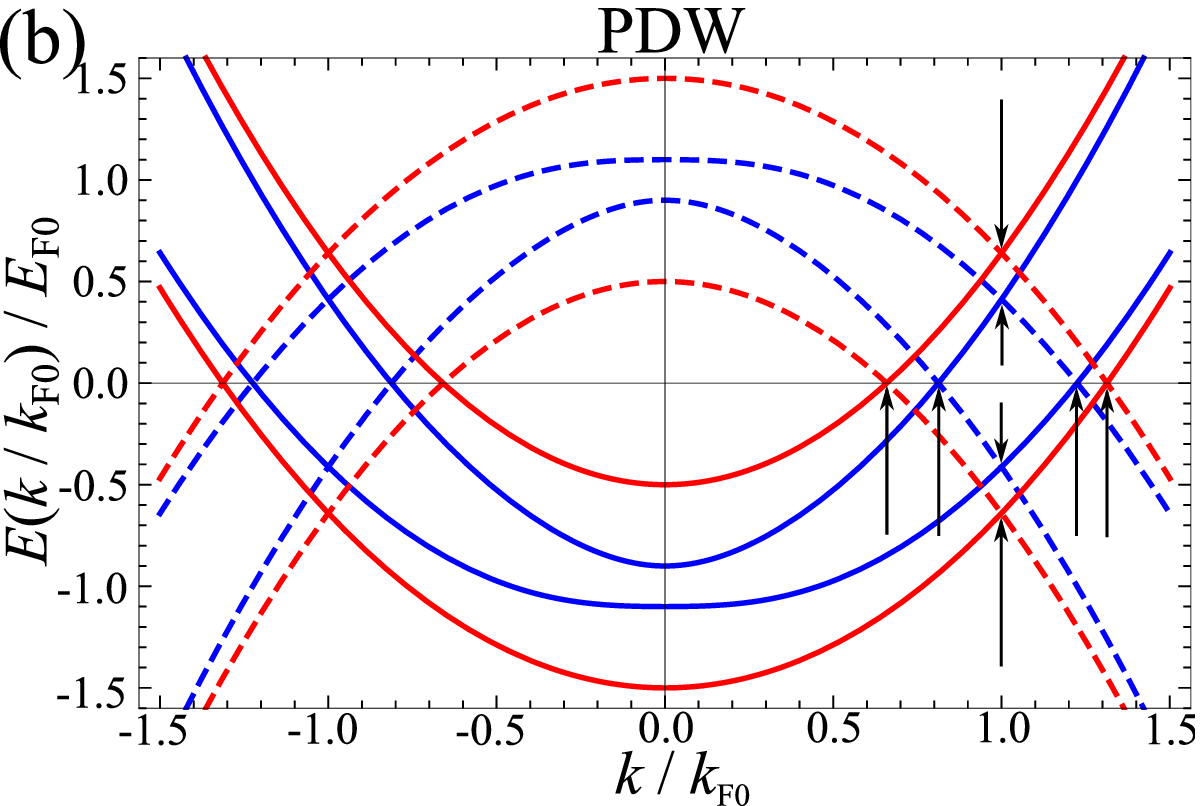}}
    \end{tabular}
\caption{
\label{fig6} (Color online)
Normal state energy spectra at
$\alpha/E_{\rm F0}=0.4$.
Arrows indicate the intersections of electron and hole bands
which form Cooper pairs in the BCS state (a) and the PDW state (b).
The horizontal axis denotes the wave number normalized by $k_{{\rm F}0}$.
Other parameters are set to $\mu_{\rm B}H/E_{\rm F0}=0.3$, $t_\perp/E_{\rm F0}=0.2$ and $\mu/E_{\rm F0}=1$
with $E_{\rm F0}=k^2_{\rm F0}/2m_{\rm e}$.
}
  \end{center}
\end{figure}

In the presence of the SOC,
the interband Cooper pairs are formed in the BCS state,
whereas the intraband Cooper pairs as well as the interband Cooper pairs are formed in the PDW state.
Figs.~\ref{fig6}(a) and \ref{fig6}(b) depict the electron (hole) band in the normal state
from bottom (top) to top (bottom) at $k=0$;
$\pm E^{\rm Low}_+(k)=\pm [\xi(k)-E_+]$,
$\pm E^{\rm Low}_-(k)=\pm [\xi(k)-E_-]$,
$\pm E^{\rm Upp}_-(k)=\pm [\xi(k)+E_-]$ and
$\pm E^{\rm Upp}_+(k)=\pm [\xi(k)+E_+]$
with $E_\pm=\sqrt{h^2_\pm+(\tilde{k}\alpha)^2}$ and $\tilde{k}=k/k_{\rm F0}$.
Unlike in Fig.~\ref{fig5},
the blue (gray) and red (light gray) lines in Figs.~\ref{fig6}
indicate the energy spectra of
subsector Hamiltonian described by
the effective magnetic fields $h_-$ and $h_+$, respectively.

For multilayered Rashba SCs
or two dimensional Rashba SCs in the presence of both the SOC and Zeeman field,
the band representation of the superconducting order parameter has been obtained
by several authors \cite{maruyama2012,alicea,shitade,masaki}.
By carrying out the unitary transformation,
the band representation of the order parameter in the BCS state is
given in the band basis as
\begin{eqnarray}
&&\hat{\varDelta}^{\rm b}_{\rm BCS}(\bm{k})=\hat{U}^\dag_k \hat{\varDelta}_{\rm BCS} \hat{U}^\ast_{-k}=
\left(
\begin{array}{cc}
0_{2\times2} & \varDelta^{\rm b-+}_{\rm BCS}(\bm{k}) \\
\varDelta^{\rm b+-}_{\rm BCS}(\bm{k}) & 0_{2\times2}
\end{array}
\right),\nonumber \\
&&\\
&&\varDelta^{\rm b-+}_{\rm BCS}(\bm{k}) =\cfrac{i\alpha \tilde{k}e^{-i\phi_k}\varDelta}{2\sqrt{E_-E_+}} \nonumber \\
&& \times
\left(
\begin{array}{cc}
-\cfrac{E_-+h_-+E_++h_+}{\sqrt{(E_-+h_-)(E_++h_+)}}
 &
\cfrac{E_-+h_--(E_+-h_+)}{\sqrt{(E_-+h_-)(E_+-h_+)}}
\\
\cfrac{-(E_--h_-)+E_++h_+}{\sqrt{(E_--h_-)(E_++h_+)}}
 & 
\cfrac{E_--h_-+E_+-h_+}{\sqrt{(E_--h_-)(E_+-h_+)}}
\end{array}
\right),\nonumber \\
&& \\
&&\varDelta^{\rm b+-}_{\rm BCS}(\bm{k}) = {i\alpha \tilde{k}e^{-i\phi_k}\varDelta\over 2\sqrt{E_-E_+}} \nonumber \\
&& \times
\left(
\begin{array}{cc}
-\cfrac{E_-+h_-+E_++h_+}{\sqrt{(E_-+h_-)(E_++h_+)}}
 &
\cfrac{-(E_--h_-)+E_++h_+}{\sqrt{(E_--h_-)(E_++h_+)}}
\\
\cfrac{E_-+h_--(E_+-h_+)}{\sqrt{(E_-+h_-)(E_+-h_+)}}
 & 
\cfrac{E_--h_-+E_+-h_+}{\sqrt{(E_--h_-)(E_+-h_+)}}
\end{array}
\right). \nonumber \\
\end{eqnarray}
Here, $\hat{U}_k$ is the unitary matrix diagonalizing $\hat{\mathscr H}_0(\bm{k})$.
Similarly, the band representation of the order parameter in the PDW state is obtained as
\begin{eqnarray}
\hat{\varDelta}^{\rm b}_{\rm PDW}(\bm{k})&=&\hat{U}^\dag_k \hat{\varDelta}_{\rm PDW} \hat{U}^\ast_{-k}=
\left(
\begin{array}{cc}
\varDelta^{\rm b-}_{\rm PDW}(\bm{k}) & 0_{2\times2} \\
0_{2\times2} & \varDelta^{\rm b+}_{\rm PDW}(\bm{k})
\end{array}
\right),\nonumber\\
&&\\
\varDelta^{\rm b-}_{\rm PDW}(\bm{k}) &=&
\cfrac{ie^{-i\phi_k}\varDelta}{E_-}
\left(
\begin{array}{cc}
-\alpha\tilde{k} & h_- \\
h_- & \alpha\tilde{k}
\end{array}
\right),\\
\varDelta^{\rm b+}_{\rm PDW}(\bm{k}) &=&
\cfrac{ie^{-i\phi_k}\varDelta}{E_+}
\left(
\begin{array}{cc}
-\alpha\tilde{k} & h_+ \\
h_+ & \alpha\tilde{k}
\end{array}
\right).
\end{eqnarray}
The spectral gap corresponding to
each component of the order parameter is indicated by arrows in Fig.~\ref{fig6}.
We notice that the superconducting gaps open at
eight intersections of electron and hole bands.
The feature is different
from that at $\alpha=0$.
Thus, the effect of the SOC on the internal structure of Cooper pairs
is clarified by using the band representation.
\begin{figure}[tb]
  \begin{center}
    \begin{tabular}{p{85mm}p{85mm}}
      \resizebox{85mm}{!}{\includegraphics{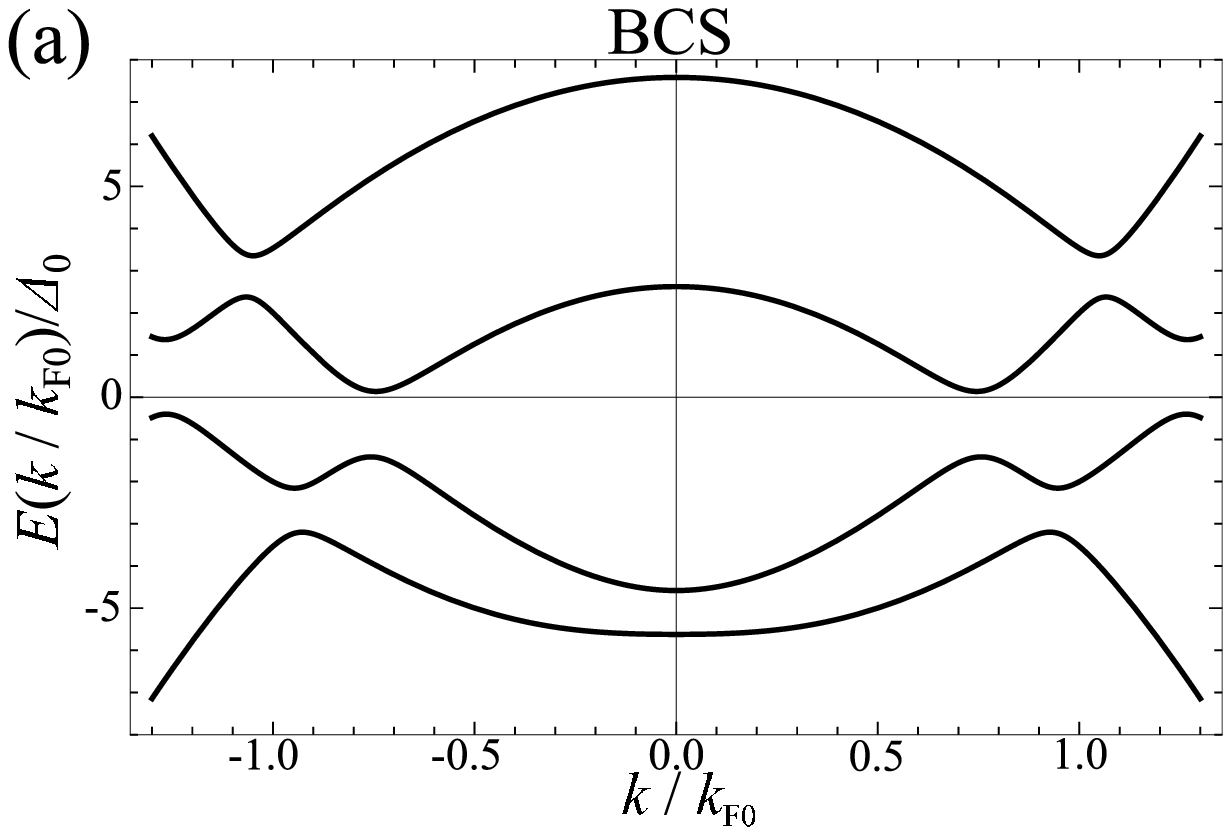}}\\
      \resizebox{85mm}{!}{\includegraphics{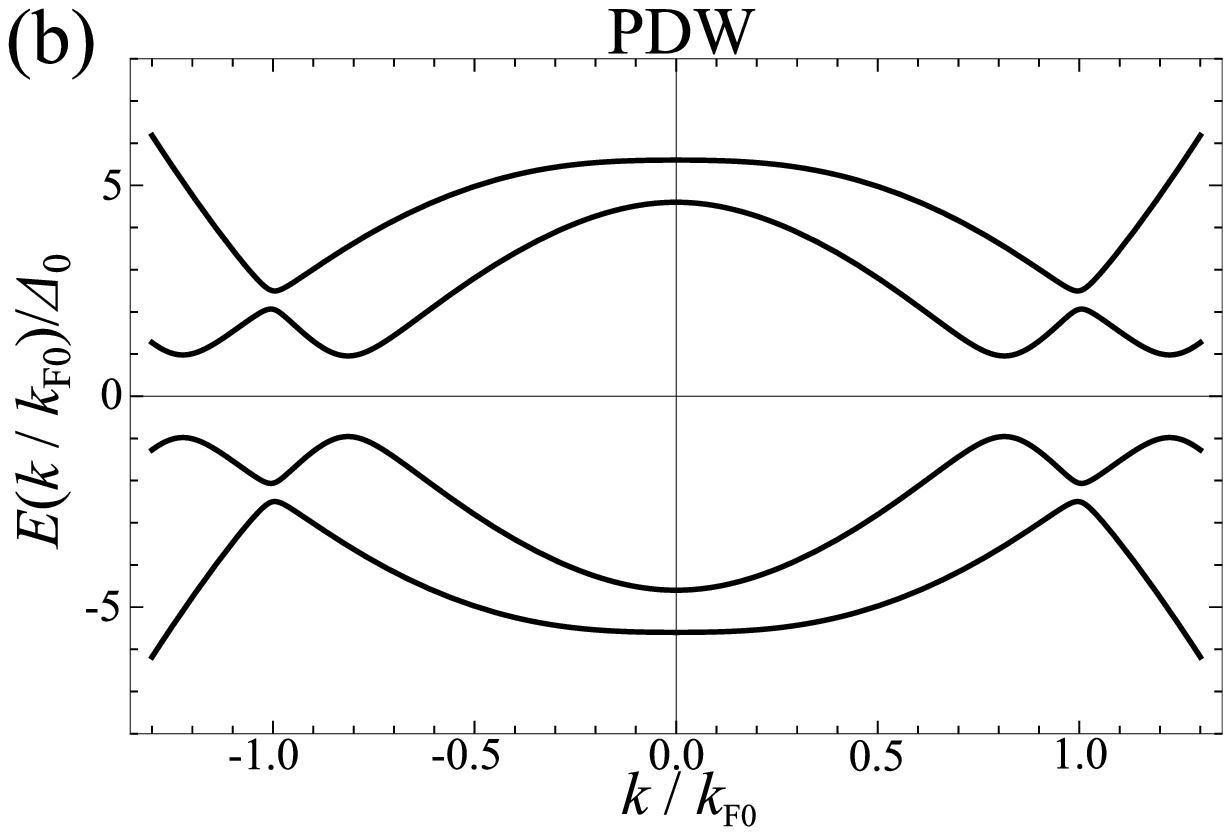}}
    \end{tabular}
\caption{
\label{fig7}
Energy spectra of the mirror subsector Hamiltonian
in the BCS state (a) and that with $h_-$ in the PDW state (b) for $\alpha/\varDelta_0=2$.
The superconducting gap energy is set to $|\varDelta|/\varDelta_0=1$.
Other parameters are set to $\mu_{\rm B}H/\varDelta_0=1.5$, $t_\perp/\varDelta_0=1$ and $E_{\rm F}/\varDelta_0=5$.
} 
  \end{center}
\end{figure}

First, in the BCS state [Fig.~\ref{fig6}(a)],
the superconducting gap does not open just at $E_{\rm F}$ [$E(k)=0$].
In order to clarify this point,
in Fig.~\ref{fig7}(a),
we show the energy spectra
of a mirror subsector
of the block-diagonalized BdG Hamiltonian in the BCS superconducting state.
We now understand that the particle-hole asymmetry in the subsector Hamiltonian leads
to the shift of the superconducting-gap center from $E_{\rm F}$.
This is because
of the even-mirror-parity of the BCS state.
The other mirror subsector
also gives the energy spectra without particle-hole symmetry.
When the superconducting gap opens in the bulk,
quasiparticle bound states (Andreev bound states) are formed
at the
core of a singly quantized vortex
around the superconducting gap center.
Thus, in the BCS state,
the
energy
of Andreev bound states at the vortex core
shifts
from $E_{\rm F}$ to a finite energy,
leading to the splitting of the ZEP
demonstrated in Fig.~\ref{fig3}(a).
%
%
%

Next, in the PDW state,
the superconducting gaps open at $E_{\rm F}$, as shown by arrows in Fig.~\ref{fig6}(b).
Because of the odd-mirror-parity in the PDW state,
particle-hole symmetry is preserved in the subsector Hamiltonian \cite{ueno2013,PhysRevLett.115.027001}.
Indeed, Fig.~\ref{fig7}(b) shows
particle-hole symmetry
in the energy spectra
for a mirror subsector of the block-diagonalized BdG Hamiltonian.
The particle-hole symmetry preserved in the subsector allows the Cooper pairs formed by quasiparticles at $E_{\rm F}$.
In Fig.~\ref{fig6}(b),
four arrows indicate the Cooper pairing at $E_{\rm F}$,
which is caused by the diagonal component of $\hat{\varDelta}^{\rm b}_{\rm PDW}(\bm{k})$.
The appearance of the diagonal components indicates the intraband Cooper pairs induced by the SOC.
Inner and outer two arrows at $E_{\rm F}$ in Fig.~\ref{fig6}(b) show
the pairing states in the mirror subsector with $h_-$ and $h_+$
[diagonal components of $\varDelta^{\rm b -}_{\rm PDW}(\bm{k})$ and $\varDelta^{\rm b +}_{\rm PDW}(\bm{k})$], respectively.
Four arrows far from $E_{\rm F}$ show
the interband pairing state
described by the off-diagonal components of $\varDelta^{\rm b -}_{\rm PDW}(\bm{k})$ and $\varDelta^{\rm b +}_{\rm PDW}(\bm{k})$.
As illustrated in the above discussion, in the PDW state
four arrows indicate the Cooper pairing at $E_{\rm F}$,
which is caused by the diagonal component of $\hat{\varDelta}^{\rm b}_{\rm PDW}(\bm{k})$.
Correspondingly, at the vortex core,
the ZEP of vortex bound states appears as already shown in Fig.~\ref{fig3}(b).

The two gap feature of the quasiparticle spectrum in Fig.~\ref{fig3}(b)
is also naturally understood by the band representation of the order parameter.
The bulk amplitudes of the intraband order parameter in subsectors
with effective magnetic field $h_-$ and $h_+$ are
$|\varDelta^{\rm b -}_{\rm PDW}(\bm{k})|=|\alpha \tilde{k} \varDelta/E_-|$ and
$|\varDelta^{\rm b +}_{\rm PDW}(\bm{k})|=|\alpha \tilde{k} \varDelta/E_+|$, respectively
[$|\varDelta^{\rm b -}_{\rm PDW}(\bm{k})| > |\varDelta^{\rm b +}_{\rm PDW}(\bm{k})|$].
Thus, the two gap-edges at low and high energies near $E_{\rm F}$ in Fig.~\ref{fig3}(b)
correspond to $|\varDelta^{\rm b +}_{\rm PDW}(\bm{k})|$ and  $|\varDelta^{\rm b -}_{\rm PDW}(\bm{k})|$, respectively.
We note that the two gap feature in Fig.~\ref{fig3}(b) does not stem from the parity mixing of the order parameter,
which is neglected in the present work.



\section{Discussions}
\label{Sec:Discussions}
\subsection{Possible realization in real crystalline materials}
We here discuss possible realization of the PDW ground state in real crystalline materials
in the presence of vortices in a high magnetic field.
In order to examine if the PDW state stabilizes in a high magnetic field,
one needs to evaluate numerically the free energy in the vortex lattice state
by employing the Brandt-Pesch-Tewort approximation or full numerical calculation
to solve quasiclassical equations.
Although we leave
the evaluation of the free energy in the vortex lattice state for a future work,
the PDW state is stable in the vortex lattice state in a situation discussed below.

Since the spatial modulation of the order parameter due to vortices occurs within $x$-$y$ plane,
this modulation does not affect seriously the phase modulation along the $z$-axis
specifying the PDW state, if the density of vortices is not large.
In heavy fermion compounds,
a large effective electron mass gives rise to a short coherence length
and a large orbital limit of upper critical field $H_{\rm c2}^{\rm orb}$.
We consider the high magnetic field region
where $H$ satisfies the condition $H^{\rm P}_{\rm c2}<H\ll H_{\rm c2}^{\rm orb}$ at low temperature,
with the conventional Pauli-limiting field $H^{\rm P}_{\rm c2}=\sqrt{2}\varDelta_0/g\mu_{\rm B}$
and the electronic $g$-factor $g$.
Thus,
the density of vortices is not large
and each vortex is sufficiently separated
from other vortices.
In this situation,
we may consider that the PDW state is
stabilized also in the vortex lattice state.
Indeed, focusing on a single vortex, which is far separated from the other vortices,
we have already showed that a self-consistent solution of the order parameter exists only in the PDW state
in high magnetic fields $\mu_{\rm B}H/T_{\rm c0}\gtrsim 1.5$ \cite{higashi2014multilayer}.
In such a high magnetic field, the vortex solution of the order parameter in the BCS state does not exist,
since the BCS state is completely suppressed by the paramagnetic depairing effect \cite{higashi2014multilayer}.
This is consistent with the result in the paramagnetic limit \cite{yoshida2012}.

In a short coherence length situation discussed above,
the dominant paramagnetic depairing effect stabilizes the PDW state in the vortex state.
Hence, we assume SCs with a large Maki parameter $\alpha_{\rm M}=\sqrt{2}H_{\rm c2}^{\rm orb}/H^{\rm P}_{\rm c2}$ \cite{matsuda-shimahara}.
This situation is often realized in heavy fermion compounds.
One of the promising candidate compounds
in which the paramagnetic depairing effect plays a dominant role in the Cooper pair destruction mechanism
is a representative heavy fermion SC CeCoIn$_5$ \cite{thompson2012}.
Thus, it is plausible that the epitaxial heavy fermion superlattice CeCoIn$_5$/YbCoIn$_5$ is a candidate compound of the PDW state.
Indeed, it shows a large value of $H^{\rm orb}_{\rm c2}$ \cite{mizukami2011,PhysRevLett.109.157006} ($i.e.,$ large $\alpha_{\rm M}$)
as in the bulk CeCoIn$_5$.


\subsection{Particle-hole symmetry in mirror subsector}
The suppression of the paramagnetic depairing effect in the PDW state
is related to particle-hole symmetry in the mirror subsector.
A subsector Hamiltonian in the bilayer PDW state
is equivalent to the Hamiltonian of two dimensional Rashba SCs \cite{PhysRevLett.115.027001},
in which the particle-hole symmetry is preserved.
In two dimensional Rashba SCs,
the paramagnetic depairing is suppressed in a Zeeman field parallel to $z$-axis when the SOC is sufficiently large.
This is an intuitive understanding of the paramagnetic deparing suppressed in the PDW state.

On the other hand, the BCS state is affected by the paramagnetic depairing effect.
The mirror subsector in the BCS state lacks particle-hole symmetry
due to the existence of an interlayer coupling and a Zeeman field.
Thus a subsector Hamiltonian is no longer the Hamiltonian of SCs.
Then, the BCS state is affected by the paramagnetic depairing effect, and the ZEP splits off.

The order parameter in a layer in both BCS and PDW states can be viewed as a uniform isotropic $s$-wave pairing
without center of mass momentum of Cooper pairs.
Thus, one can just seek the conventional vortex solution
without the Fulde-Ferrell-Larkin-Ovchinnikov (FFLO) modulation within the layer by self-consistent calculations.
The FFLO vortex core does not necessarily possess the zero energy states unlike the PDW vortex core.
As shown in Refs. \cite{mizushima2005} and \cite{ichioka2007},
the zero energy LDOS at the vortex core split into two peaks
at finite energies due to the Zeeman field,
and each peak corresponds to a spin component.
It is possible to distinguish the PDW vortex core from the FFLO one
by investigating the presence or absence of
zero-energy quasiparticle bound states at a vortex center.
We again stress that unusual spectral features in the PDW state comes from the SOC and particle-hole symmetry of the mirror subsector.

\subsection{Sudden vortex core shrinkage}
As shown in Fig.~\ref{fig7}(b),
a large bulk spectral gap $\Delta E(k)$ opens at $E_{\rm F}$ in the PDW state,
while there appears a small energy gap in the BCS state [Fig.~\ref{fig7}(a)]
(note that Fig.~\ref{fig7} depicts the energy spectra for only one subsector Hamiltonian).
This results from the particle-hole symmetry (asymmetry) in the mirror subsector in the PDW (BCS) state.
The vortex core radius is characterized by the coherence length,
which is inversely proportional to the magnitude of $\Delta E(k)$.
In the BCS state, the small $\Delta E(k)$ gives rise to the large vortex core size.
On the other hand,
the large $\Delta E(k)$ appears due to the change of the internal structure of the Cooper pair by increasing the magnetic field through the BCS-PDW transition,
leading to a sudden vortex core shrinkage.

A possible manifestation of the BCS-PDW phase transition may be observed in an entropy jump,
giving rise to the increase of superlattice temperature as the latent heat
through the BCS-PDW first order transition \cite{yoshida2012}
with increasing a magnetic field.
In general, it is difficult to observe bulk quantities in thin films.
However, the site-selective NMR experiment
has succeeded in epitaxial superlattices CeCoIn$_5$/YbCoIn$_5$ \cite{yamanaka2015}.
Thus, in the vicinity of the BCS-PDW transition field,
the zero energy DOS obtained from the NMR spectra might depend on a magnetic field sublinearly,
reflecting the decrease of low-energy excitations due to the vortex core shrinkage.
This change in the low-energy excitations may occur not only in the isotropic $s$-wave state
studied in this paper but also in the $d$-wave pairing state expected in the superlattices CeCoIn$_5$/YbCoIn$_5$.
Direct observation of the vortex core shrinkage by STM/STS is also promising way to detect the BCS-PDW transition.


\section{Summary}
\label{Sec:VI}
We have numerically investigated the electronic structure of a vortex core in bilayer Rashba SCs
by means of the self-consistent quasiclassical calculation.
We found that the LDOS structure in the PDW state is quite different from that in the BCS state.
The zero energy vortex bound state exists in the PDW state, whereas it is absent in the BCS state due to the Zeeman effect.
This prominent difference stems from
(i) the presence or absence of particle-hole symmetry in the mirror subsector of the block-diagonalized BdG Hamiltonian
and
(ii) the internal structure of the Cooper pair influenced by the SOC.
Another intriguing feature of the PDW state is the small vortex core size compared with the BCS state, leading to a sudden shrinkage of vortex cores through 
the BCS-PDW phase transition.
The characteristic vortex core structure in the PDW state may be observed by STM/STS and/or NMR measurements at low temperatures.
The exotic superconducting phase under a magnetic field may be identified by investigating these features.

\section*{Acknowledgments}
We thank Masaru Kato for fruitful discussions and reading this manuscript.
We also thank Takuto Kawakami, Noriyuki Kurosawa, Tuson Park and Yuji Matsuda for helpful discussions or comments.
The computation in this work has been done using the facilities
of the Supercomputer Center of the Institute for Solid State Physics, the University of Tokyo.
T. Y. is supported by a JSPS Fellowship for Young Scientists.
This work was supported by the ``Topological Quantum Phenomena" (No. 25103711) and ``J-Physics" (15H05884) Grant-in Aid for Scientific Research on Innovative Areas from MEXT of Japan, and by JSPS KAKENHI Grant Numbers 24740230, 15K05164, and 15H05745.


\end{document}